# Implementation of tangent linear and adjoint models for neural networks based on a compiler library tool

Sa Xiao,[a,b]    Hao Jing*,[a,b]    Honglu Sun,[a,b]    Haoyu Li,[a,b]

[a] *State Key Laboratory of Severe Weather Meteorological Science and Technology (LaSW), Beijing, China*

[b] *CMA Earth System Modeling and Prediction Centre (CEMC), Beijing, China*

**Abstract**

This paper presents TorchNWP, a compilation library tool for the efficient coupling of artificial intelligence components and traditional numerical models. It aims to address the issues of poor cross-language compatibility, insufficient coupling flexibility, and low data transfer efficiency between operational numerical models developed in Fortran and Python-based deep learning frameworks. The tool consists of standalone, pluggable, and uniformly configurable compilation and runtime scripts. Based on LibTorch, it optimizes and designs a unified application-layer calling interface, converts deep learning models under the PyTorch framework into a static binary format, and provides C/C++ interfaces. Then, using hybrid Fortran/C/C++ programming, it enables the deployment of deep learning models within numerical models. Integrating TorchNWP into a numerical model only requires compiling it into a callable link library and linking it during the compilation and linking phase to generate the executable. On this basis, tangent linear and adjoint model based on neural networks are implemented at the C/C++ level, which can shield the internal structure of neural network models and simplify the construction process of four-dimensional variational data assimilation systems. Meanwhile, it supports deployment on heterogeneous platforms, is compatible with mainstream neural network models, and enables mapping of different parallel granularities and efficient parallel execution. Using this tool requires minimal code modifications to the original numerical model, thus reducing coupling costs. It can be efficiently integrated into numerical weather prediction models such as CMA-GFS and MCV, and has been applied to the coupling of deep learning-based physical parameterization schemes (e.g., radiation, non-orographic gravity wave drag) and the development of their tangent linear and adjoint models, significantly improving the accuracy and efficiency of numerical weather prediction.

## 1. Introduction

There are two technical routes for artificial intelligence to empower weather forecasting. One takes numerical weather prediction (NWP) as the core, and uses artificial intelligence techniques to improve the accuracy and computational efficiency of the whole workflow, including observation processing, data assimilation, model prediction, and product interpretation. The other takes artificial intelligence as the core, and establishes data-driven meteorological forecasting models that integrate physical mechanisms. For the first routine, traditional numerical weather prediction models are mainly developed in Fortran as the standard programming language. However, the core components of artificial intelligence models are mostly trained or deployed using machine learning frameworks based on the dynamically typed language Python. This means that AI models cannot be directly invoked as functions by numerical weather prediction models written in the statically typed language Fortran.

Traditional numerical models and machine learning models are independent modules with their own internal data structures, data distributions, and parallel decompositions. It is desirable to achieve flexible data transmission between them without introducing extra dependencies and excessive code modifications to the originally intact program, while supporting the functions of full variable-field data restructuring and fine-grained data exchange, which addresses the issue of data transmission flexibility. To couple deep learning models with traditional numerical models, one approach is to export parameters (such as weights and biases) from the trained deep learning models and embed them into Fortran code in the form of hard coding. However, this approach suffers from low development efficiency: any modification to the parameters of the deep learning model requires corresponding changes to the relevant code. Meanwhile, the lack of mature and user-friendly machine learning frameworks and tools in the Fortran ecosystem results in a very high cost for porting deep learning models.

One approach is the artificial intelligence model interface similar to the Fortran-Keras Deep Learning Bridge (Ott et al., 2020), which can import trained deep learning parameters from outside the Fortran program, store them in TXT files, and call them during inference and prediction of the deep learning model. The Fortran-based implementation ensures relatively flexible deployment in numerical models, allowing model replacement by simply modifying the parameters in the TXT files without altering the network structure of the deep learning model, yet this interface mainly supports deep learning models based on fully connected layers and cannot support many state-of-the-art network architectures such as Stable Diffusion or Transformer models.

Another commonly used coupling approach separates the Fortran-based host model and Python-based deep learning components into two groups of processes, and achieves data transmission through inter-process communication (IPC) with synchronization control during coupled running (Wang et al., 2022). During coupled execution, the IPC data interface receives input variables from the host model, performs gathering operations on the raw data to improve data transmission efficiency, then carries out floating-point endian conversion and sends the data to the artificial intelligence model via IPC mechanisms. The IPC data interface supports multiple communication methods, including MPI, FIFO pipes, and high-speed data buffers. After the artificial intelligence inference engine completes computation, it first writes the data back to the IPC data interface, and then returns the results to the host model through data gathering and floating-point endian conversion. Throughout the entire data coupling process, a synchronization controller governs the operational synchronization between the host model and the artificial intelligence inference engine.

Fortran Torch Adapter (FTA) is a general-purpose tool based on LibTorch that provides users with an interface for directly manipulating neural network models in Fortran via the FTA library (Mu et al., 2023). This tool supports converting deep learning models under the PyTorch framework into a static binary format and provides C/C++ interfaces; then, using hybrid Fortran and C/C++ programming techniques, it deploys deep learning models within numerical models. Compared with other coupling methods, FTA offers higher flexibility, but certain code compatibility issues still exist during hybrid programming. Meanwhile, since LibTorch follows C/C++ programming conventions, debugging and stability testing during coupled execution remain insufficiently flexible and efficient, and the limited functionality of FTA cannot meet the full requirements for the integration of artificial intelligence and numerical weather prediction.

The coupling of artificial intelligence models and numerical models typically involves frequent invocations of the same network for fast inference and intensive communication, which is suitable for real-time linking using LibTorch-based compilation library tools to ensure efficient runtime execution of the hybrid model. Therefore, developing a comprehensive coupling middleware that supports heterogeneous platform deployment, is compatible with mainstream neural network models, and enables mapping of different parallel granularities as well as efficient parallel execution represents a core requirement for achieving the deep integration of artificial intelligence and numerical weather prediction.

## 2. Introduction to TorchNWP

To address technical challenges in the coupling process between artificial intelligence inference modules and numerical models, including inconsistent cross-language interface designs, incompatible operation modes, and non-uniform data transmission methods, we have developed TorchNWP -- a fully functional, highly efficient, stable, and adaptable compilation library tool for the effective coupling of artificial intelligence components with traditional numerical models. Based on standardized interface design, cross-language binding, efficient model loading and inference, comprehensive performance optimization, and rigorous testing and validation, TorchNWP enables low-cost, high-efficiency integration of deep learning models with traditional numerical weather prediction models. Meanwhile, through the standardized construction of tangent linear and adjoint models, it simplifies the development of four-dimensional variational data assimilation systems, provides core support for the deep integration of artificial intelligence and numerical weather prediction technologies, and helps improve the accuracy and computational efficiency of numerical weather prediction. The structural flowchart of this compilation library tool is shown in Figure 1. Detailed descriptions are given as follows:

(1) Interface Design. The interface design is developed based on C++-based TorchScript, with the core objective of building a concise, unified, and efficient interaction bridge between Fortran numerical models and PyTorch deep learning models, shielding underlying technical complexities and reducing the difficulty of user invocation. Specifically, by explicitly defining three core interface functions at the C++ layer, the full-process interaction between Fortran and PyTorch models is realized. The functions and roles of each function are as follows:

**The model_new function:** It is mainly responsible for model initialization, including allocating memory resources required for model operation, loading model configuration parameters, and initializing the LibTorch runtime environment, laying the groundwork for subsequent model loading and inference. This function supports receiving user-configured parameters such as model storage paths and runtime devices (CPU/GPU), featuring good flexibility to adapt to model initialization requirements in different scenarios.

**The model_forward function:** As the core interface for model inference, it is responsible for receiving input data transmitted from the Fortran numerical model, invoking the model to complete inference calculations, and returning the inference results to the Fortran numerical model. Built-in auxiliary functions such as data format verification and dimension conversion are integrated within this function to ensure that input data meets the requirements of model inference, while guaranteeing the accuracy and completeness of inference results.

**The model_delete function:** It is primarily responsible for resource release after model operation, including freeing up memory resources occupied by the model, destroying the LibTorch runtime environment, and clearing input and output data caches. This prevents memory leakage issues, ensures the long-term stable operation of the tool, and adapts to the operational requirements of numerical models for long-term continuous integration.

The above three core interface functions work in coordination to form a complete link for Fortran numerical models to call PyTorch models. Users do not need to focus on the implementation details of the underlying C++ and LibTorch; they only need to call the corresponding interface functions to complete the initialization, inference, and resource release of deep learning models, which significantly reduces the difficulty of cross-language coupling.

(2) Binding of Fortran and C++. Due to the inherent language barriers between traditional numerical models developed in Fortran and deep learning model interfaces encapsulated in C++, including differences in data types, memory layouts, and function call specifications, this invention adopts Fortran/C/C++ mixed programming technology and leverages the ISO_C_Binding module to achieve deep binding between Fortran and C++. This thoroughly resolves cross-language compatibility issues and enables efficient data transmission and function calls between the two languages. As a standardized interface provided by the Fortran language, the ISO_C_Binding module defines clear correspondences between various data types in Fortran and C++. It can accurately convert basic data types (e.g., integers, floating-point numbers, logical values), array types, and pointer types in Fortran into their corresponding types in C++, eliminating issues such as call failures and data corruption caused by incompatible data types. For instance, it converts the REAL(8) type in Fortran to the double type in C++, and two-dimensional arrays in Fortran to pointer arrays in C++, ensuring correct data transmission between the two languages.

(3) Model Loading and Inference. In terms of model loading, this tool leverages the torch::jit::load function provided by LibTorch to load pre-trained and exported deep learning models (typically stored in .pt file format) developed under the PyTorch framework. The .pt file fully preserves all structural information of the model (including the number of network layers, the number of neurons per layer, convolution kernel parameters, activation function types, etc.) and parameter information (including weights, biases, normalization parameters, etc.). The torch::jit::load function can accurately read the information in this file, reconstruct the model structure, load model parameters, and complete the initial loading of the model. This loading method supports various mainstream PyTorch models, including Convolutional Neural Networks (CNNs), Transformers, Diffusion models, Residual Neural Networks (ResNets), etc., adapting to various application scenarios in the field of numerical weather prediction. It also features fast loading speed and high stability, meeting the real-time operation requirements of numerical models.

In terms of model inference, after the model is loaded, the module.forward function is called to execute inference calculations on the input data. As a standardized inference interface provided by the LibTorch framework, the module.forward function can receive input data that has undergone format conversion, execute the calculation operations of each layer in sequence according to the model's network structure, and finally output the inference results. This function supports batch inference, enabling efficient processing of large-scale input data transmitted by numerical models and improving inference efficiency.

It is important to note that due to the fundamental differences in array memory layout between Fortran and C++ (Fortran uses column-major storage, while C++ uses row-major storage), directly passing input data from Fortran to the model for inference will lead to data corruption and compromise the accuracy of inference results. Therefore, during the inference process, this tool adjusts the dimensions of the input data to adapt to Fortran's memory layout, ensuring that the input data can be correctly recognized and processed by the model. Specifically, by adjusting the order of data dimensions, arrays stored in Fortran's column-major format are converted into the row-major format recognizable by the model. This avoids inference errors caused by differences in memory layout while minimizing performance loss during data adjustment.

The entire process of model loading and inference is encapsulated within C++ interface functions. Fortran users do not need to focus on underlying implementation details, they only need to pass input data and obtain inference results through the interface functions to complete the invocation of deep learning models. This significantly reduces the difficulty of coupling and improves coupling efficiency.

(4) Performance Optimization. To address issues such as low data transmission efficiency, excessive memory usage, and slow runtime speed during the coupling of deep learning models and numerical models, and to ensure that the tool can meet the demands of long-term continuous integration and large-scale data processing in numerical models, this tool adopts a variety of targeted measures for performance optimization to comprehensively improve its operating efficiency and stability. The specific optimization strategies are as follows:

First, to reduce the performance overhead caused by memory layout differences between Fortran and C++, the permute function in PyTorch is used to adjust the dimensions of input and output data, enabling adaptive conversion of data layouts and avoiding unnecessary data copies. The permute function can rapidly rearrange the dimension order, converting Fortran's column-major stored data into C++'s row-major format or vice versa. The entire process only modifies index mappings without additional data copying, which significantly reduces performance overhead during data transmission and improves data transfer efficiency.

Second, the tool supports multi-device execution. Users can flexibly choose to run the deep learning model on either CPU or GPU according to actual computing resources, achieving rational allocation of computational resources and accelerating model inference. When GPU resources are available, the tool automatically enables GPU-accelerated inference, leveraging the parallel computing power of GPUs to greatly improve inference efficiency for large-scale data. When no GPU is available, the tool automatically switches to CPU mode to ensure normal operation.

In addition, this tool adopts an instant memory release mechanism that immediately frees memory occupied by input/output data and intermediate computational results after inference is completed, preventing memory leaks and excessive memory consumption. To handle the large data volumes typical in numerical weather prediction, a blocked data processing technique is used to split large-scale data into small chunks, which are loaded and inferred in batches to avoid runtime lag and memory overflow caused by excessively large single-batch data. Meanwhile, the storage scheme of model parameters is optimized using efficient compression algorithms, which reduce storage and memory overhead without compromising inference accuracy. These strategies jointly improve the overall performance of the tool.

(5) Standardized Construction of Tangent Linear and Adjoint Models. Tangent linear and adjoint technologies based on neural networks are implemented at the C++ layer to shield the impact of the internal structure of neural network models and provide standardized tangent linear and adjoint interfaces. The specific implementation process is as follows:

**Forward Calculation:** Physical quantities input by the numerical model (including various physical parameters in the numerical weather prediction process, such as temperature, humidity, wind speed, etc.) are converted through the application interface layer and then passed to the deep learning model encapsulated at the C/C++ layer to complete the forward inference process and obtain model output results.

**Tangent Linear Calculation:** Using the automatic differentiation tool of LibTorch, all differentiable operations (such as convolution, matrix multiplication, activation functions, etc.) in the forward calculation process are tracked, and the Jacobian matrix of the deep learning model's output with respect to the input is calculated to generate the tangent linear model. The tangent linear model satisfies $y' = J \cdot x'$, where $x'$ is the input perturbation, $y'$ is the output perturbation, and $J$ is the Jacobian matrix, which can accurately characterize the impact of input perturbations on output results.

**Adjoint Calculation:** A scalar loss function $L = y \cdot y^*$ is constructed (where y is the forward output result of the model and $y^*$ is the adjoint variable). Using the automatic differentiation tool of LibTorch, the gradient of the scalar loss function $L$ with respect to the input physical quantities is calculated to obtain the adjoint model result $x^* = J^T \cdot y^*$ (where $J^T$ is the transpose of the Jacobian matrix $J$).

**Interface Encapsulation:** The above tangent linear calculation process and adjoint calculation process are encapsulated into standardized C/C++ interfaces, which are bound to the Fortran language through the ISO_C_Binding module. This enables Fortran numerical models to directly call these interfaces, shielding differences in the internal structures of neural networks.

In addition, it is necessary to verify the consistency between the calculated derivatives of neural network parameters obtained by the automatic differentiation algorithm and the gradients of the forward model, as well as the consistency between the machine learning adjoint model constructed by automatic differentiation and the traditional gradient method based on adjoint models. Through a rigorous verification process, the correctness and reliability of the tangent linear and adjoint models are ensured, providing strong support for the construction of the four-dimensional variational data assimilation system.

(6) Testing and Validation. Strict testing and validation have been conducted during the development of this tool, covering functional testing, performance testing, compatibility testing, and other aspects. Tests have been carried out on various Fortran codes and PyTorch models to ensure its stability and efficiency under diverse scenarios.

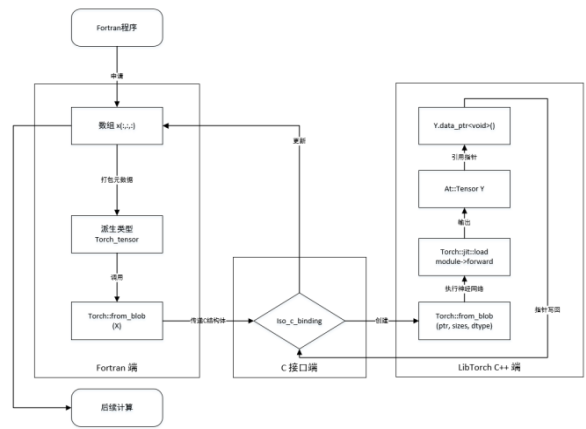

Figure 1. Structural Flowchart of the TorchNWP Compiler Library Tool.

## 3. Construction of neural network-based tangent linear and adjoint models using TorchNWP

This section uses a concrete example to illustrate the implementation and operational workflow of TorchNWP. The CMA Global Forecast System (CMA-GFS) developed by the China Meteorological Administration is used as the application platform. As a national-level core operational numerical weather prediction model, CMA-GFS runs operationally four times per day and provides high-precision 10-day global weather forecasts. As one of its core physical processes, the non-orographic gravity wave drag module consumes a significant portion of the computational time in physical process simulations. We couple the well-trained high-precision neural network parameterization scheme for non-orographic gravity wave drag into the CMA-GFS model using the TorchNWP library. Meanwhile, tangent linear and adjoint models corresponding to the neural network are constructed based on TorchNWP to support the four-dimensional variational data assimilation system. Detailed implementation is described as follows:

### 3.1 Training and coupling implementation of the high-precision neural network model for non-orographic gravity wave drag

First, based on the physical mechanism of non-orographic gravity wave drag, the input and output physical quantities for the deep learning model are selected, and a high-quality training dataset covering different seasons and weather situations is constructed. This aims to enable the deep learning model to accurately identify and learn the nonlinear characteristics, spatiotemporal evolution laws, and complex influencing factors in the non-orographic gravity wave drag process, laying a data foundation for subsequent model training and numerical model coupling. Specifically, the input physical quantities include 11 core meteorological parameters such as temperature, humidity, wind speed, pressure, geopotential height, and latitude; the output physical quantities include 5 key physical quantities related to non-orographic gravity wave drag. The total duration of the dataset is one year. Model integration is performed using a cold-start approach with 10-day integration per case and hourly data output, resulting in a total data volume of 41 TB. All data have undergone strict quality control, outlier removal, missing value imputation, and standardized preprocessing to guarantee data accuracy and integrity.

Considering the operational characteristics of numerical models, the deep-learning-based non-orographic gravity wave drag parameterization scheme adopts a design philosophy of independent processing for single vertical grid columns. That is, the model takes a single vertical grid column from the CMA-GFS model as the basic processing unit, and both input and output data are one-dimensional vectors composed of multi-physical variables. This design can perfectly match the grid structure and computational logic of the original numerical model, significantly reducing the coupling difficulty between the neural network model and the numerical model, and improving operational stability after coupling. Accordingly, the input and output data of the deep learning model are two-dimensional arrays with sizes of (11, 89) and (5, 89), respectively, where the first dimension corresponds to the number of input/output physical variables and the second dimension corresponds to the number of vertical model levels. It should be particularly noted that latitude has a significant influence on the simulation accuracy of non-orographic gravity wave drag. Therefore, latitude is included in the model as a core input variable to ensure that the model can adapt to the characteristics of non-orographic gravity wave drag in different latitude regions.

In the data preprocessing stage, the choice of normalization method directly affects model training performance and inference accuracy. After multiple rounds of comparative experiments, the optimal normalization strategy was determined as follows. For input variables, a per-variable per-level normalization method is adopted. That is, normalization is performed separately for each input physical variable and each vertical model level. This eliminates magnitude differences across variables and vertical levels and avoids the dominance of certain variables or levels during model training. For output variables, per-variable global normalization is used, meaning each output physical variable is normalized globally and uniformly to ensure the consistency and rationality of the output results. The advantage of this normalization strategy is that it accurately adapts to the significant differences in physical characteristics at different altitude levels during non-orographic gravity wave drag processes, effectively improves the model's ability to capture weak signals, and reduces systematic bias.

The architecture of the deep learning model adopts an optimized residual neural network (ResNet), to minimize the number of model parameters and computational overhead while ensuring inference accuracy, so as to meet the demand of long-term continuous integration in numerical models. By using one-dimensional convolutional neural networks, the model can efficiently extract the spatiotemporal features of physical quantities in the vertical direction and capture the vertical evolution of the non-orographic gravity wave drag process; the last few layers of the network are fully connected layers for deeply mining the nonlinear correlations among different physical variables, thus improving the model's fitting and generalization capabilities. To further enhance coupling efficiency, the normalization and denormalization operations for input and output data are integrated internally into the deep learning model, eliminating the need for additional code implementation in the numerical model and simplifying the coupling workflow.

Experimental results show that, increasing the number of parameters in the deep learning model has a significant positive effect on improving prediction accuracy; however, a larger number of parameters also leads to slower inference speed. Taking both prediction accuracy and inference speed into comprehensive consideration, after multiple rounds of optimization on the network structure, the current optimal network consists of two residual modules with 16 output channels, where each module contains two one-dimensional convolutional layers, followed by an additional one-dimensional convolutional layer that reduces the number of channels from 16 to 5 after the second residual module. Since the model output consists of three vectors and two scalars, two parallel fully connected layers are attached after the convolutional layers to perform predictions for the vectors and scalars separately.

The neural network model for non-orographic gravity wave drag was trained and optimized offline. On the test set, the deviations between the non-orographic gravity wave drag-related physical quantities predicted by the neural network and the reference values output by the model were minimal, indicating that the trained surrogate model can accurately fit the non-orographic gravity wave drag process. Based on the TorchNWP compilation library tool, the above-trained deep learning model was coupled to the original numerical weather prediction model CMA-GFS. First, the trained PyTorch model was saved as a .pt file, which fully preserves all structural information, parameter information, and normalization configurations of the deep learning model to ensure lossless accuracy during model conversion and invocation. Second, the .pt file was compiled and converted using the compilation scripts provided by the TorchNWP library to generate a C++ link library that can be called by the Fortran model; during compilation, parameters such as compiler type, PyTorch version, and runtime device (CPU/GPU) can be flexibly configured to adapt to the operational environment of the CMA-GFS model. Third, when compiling and linking the CMA-GFS model to generate an executable program, the above-generated C++ link library was linked with the original model; meanwhile, a small amount of interface calling code was written at the positions in the original model where the deep learning model needs to be enabled to invoke the three core interfaces (model_new, model_forward, model_delete) provided by TorchNWP, completing model initialization, inference, and resource release.

Long-term continuous integration tests and accuracy verification were conducted on the coupled hybrid model. The results show that after 6 hours of integration, compared with the original numerical model, the hybrid model yields a mean absolute error (MAE) of 0.055 m/s for 10-m wind speed, 0.027 m/s for 500 hPa wind speed, and 0.013 K for 500 hPa temperature. The detailed evaluation of the forecast performance is shown in Figure 2. The hybrid model can stably complete 10-day continuous integration without memory leaks, runtime lag, or solution divergence, thus meeting the operational requirements of numerical weather prediction. In terms of computational speed, both the original model and the deep learning model are currently executed on CPUs; when the parallel scale is below 8000 cores, the inference speed of the deep learning model outperforms the original non-orographic gravity wave drag physical scheme, with a twofold improvement in computational efficiency. However, at larger parallel scales, the deep learning version exhibits a reverse speedup anomaly, which can be avoided by offloading the deep learning components to GPU execution. During the entire coupling procedure, the numerical model and the artificial intelligence inference module compute within the same process without additional inter-process communication, which completely eliminates the overhead caused by inter-process communication and further enhances coupling efficiency and runtime stability.

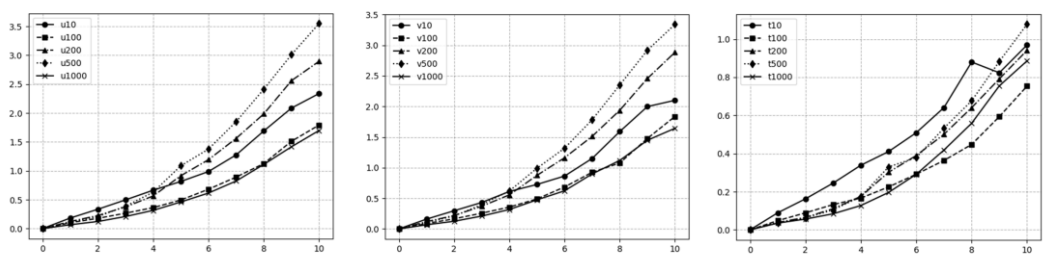

Figure 2. Evaluations of forecast performance after coupling with the high-precision neural network model for non-orographic gravity Wave Drag.

## 3.2 Implementation of Tangent Linear and Adjoint Modes for Neural Networks

After coupling the neural network model with the numerical model, to adapt to the four-dimensional variational data assimilation system, it is necessary to construct tangent linear and adjoint models corresponding to the neural network model. When traditional methods are used to build tangent linear and adjoint models, it is required to deeply understand the internal structure of the neural network and manually derive tangent linear relationships and adjoint equations. This approach not only involves an enormous workload and cumbersome processes but also is highly prone to derivation errors. Meanwhile, it cannot adapt to neural network models with different structures, resulting in poor generality. In this invention, by implementing neural network-based tangent linear and adjoint technologies at the C/C++ layer and relying on the automatic differentiation tool of LibTorch to shield the impact of the internal structure of neural network models, standardized tangent linear and adjoint models can be constructed quickly and accurately.

Let x denote the input, y the output, and M the neural network, then the tangent linear model is given by $\Delta y = J \cdot \Delta x$, where $\Delta x$ is the input perturbation, and $\Delta y$ is the output perturbation.

$$J = \begin{pmatrix} \frac{\partial y_1}{\partial x_1} & ... & \frac{\partial y_1}{\partial x_n} \\ ... & ... & ... \\ \frac{\partial y_m}{\partial x_1} & ... & \frac{\partial y_m}{\partial x_n} \end{pmatrix}$$

represents the Jacobian matrix of the neural network with respect to the input. For the adjoint model, the adjoint model at point x can be defined as $\left(\frac{\partial M(x)}{\partial x}|_{x=x_0}\right)^T$, and the following relation also holds:

$$\frac{\partial k}{\partial y} = z$$

$$\frac{\partial k}{\partial x_0} = \left(\frac{\partial y}{\partial x_0}\right)^T \frac{\partial k}{\partial y} = \left(\frac{\partial M(x)}{\partial x}|_{x=x_0}\right)^T z,$$

Combined with the computation graph in Figure 3, given the adjoint at point y is known as z, the inference state y is obtained by feeding $x_0$ into the neural network model M. A scalar k is then constructed via the dot product between y and z. The gradient of the scalar k with respect to $x_0$ gives the adjoint result at point $x_0$.

Based on the above derivation, in Python, the calculation of the Jacobian matrix and the gradient of k with respect to $x_0$ can both be implemented using the automatic differentiation tool of PyTorch. The fundamental reason is that a neural network is a composite function composed of a series of differentiable operations (such as convolution, activation functions, and matrix multiplication), and automatic differentiation can accurately track the chain rule-based derivative calculation process of these operations. While avoiding numerical differentiation errors and redundant computations in symbolic differentiation, it efficiently outputs the Jacobian matrix (full partial derivatives for multi-input multi-output scenarios) and gradients (partial derivatives of a single output with respect to multiple inputs), which caters to the derivative calculation requirements of neural networks. Meanwhile, this also indicates that the mechanism of calculating gradients via automatic differentiation is essentially identical to the computational scheme of traditional tangent linear and adjoint models.

Considering the need for coupling with a Fortran-based four-dimensional variational data assimilation (4D-Var) system, we leverage compiler library tools. By using the C++ version of Torch, we implement automatic differentiation functionality in the coupling middle layer via C++, thereby realizing a neural network-based tangent linear and adjoint model. Since the model information is fully preserved in .pt files, this implementation scheme for the tangent linear and adjoint models can shield the impact of the internal model structure and simplify the construction process of the four-dimensional variational data assimilation system.

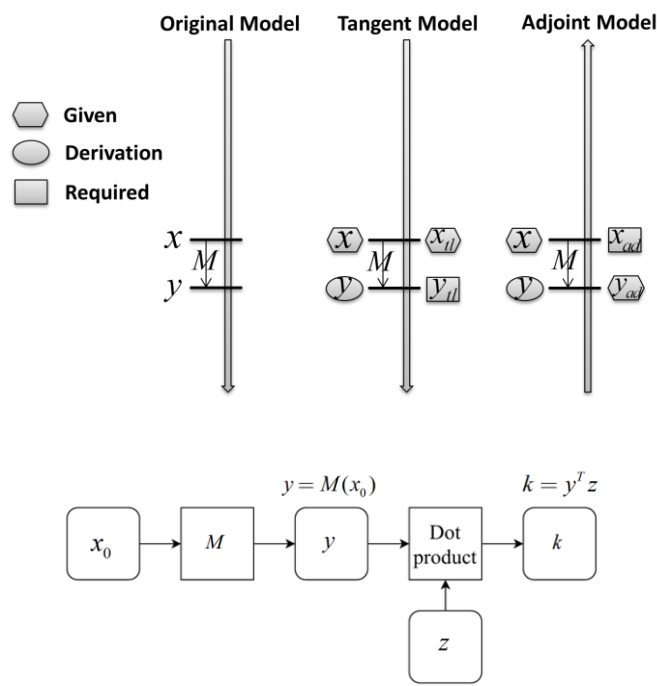

Figure 3. Computational graph of the tangent linear and adjoint model.

4. **Conclusion**

TorchNWP has been successfully integrated into national-level numerical weather prediction (NWP) models such as CMA-GFS, supporting the coupling of deep learning parameterization schemes for multiple physical processes including radiation and non-orographic gravity wave drag. Its core advantages are summarized as follows:

(1) Compatibility: It resolves the cross-language compatibility issue between Fortran-based numerical models and Python-based deep learning frameworks, supports mainstream deep learning models such as residual networks (ResNets) and Transformers, and is compatible with CPU/GPU heterogeneous platforms;

(2) Efficiency: Model invocation is completed within the same process, eliminating communication overhead and achieving high data transmission efficiency and fast model inference speed;

(3) Low invasiveness: Only compilation/linking and a small amount of interface calling code are required, without the need for substantial modifications to the original numerical model, thus reducing the coupling cost;

(4) Scalability: It supports flexible configuration of compilers, model parameters, and parallel granularity, adapting to the customized requirements of different numerical models;

(5) Assimilation adaptability: It standardizes the interfaces for tangent linear/adjoint models, shields the black-box characteristics of neural networks, simplifies the construction process of four-dimensional variational data assimilation (4D-Var) systems, and reduces the cost of assimilation adaptation.

**Acknowledgment**

This work was supported by the National Natural Science Foundation of China (Grant No. U2342220). We also thank Dr. Zhang Lin and Dr. Liu Yongzhu for their discussions.